\documentstyle[aps,prb,floats,epsfig,twocolumn]{revtex}

\begin{document}
\draft
\wideabs{
\title{Role of the dimerized gap due to anion ordering in spin-density wave phase of
(TMTSF)$_2$ClO$_4$ at high magnetic fields}
\author{N.~Matsunaga\cite{E-mail}}
\address{Division of Physics, Hokkaido University, Sapporo 060-0810, Japan\\
and CRTBT-CNRS, laboratoire associ\'{e} \`{a} l'UJF, BP 166, 38042 Grenoble Cedex 9, France\\
and Grenoble High Field Laboratory, CNRS/MPI-FKF, BP 166, 38042 Grenoble Cedex 9, France}
\author{A.~Ayari and P.~Monceau}
\address{CRTBT-CNRS, laboratoire associ\'{e} \`{a} l'UJF, BP 166, 38042 Grenoble Cedex 9, France\\
and Grenoble High Field Laboratory, CNRS/MPI-FKF, BP 166, 38042 Grenoble Cedex 9, France}
\author{A.~Ishikawa and K.~Nomura}
\address{Division of Physics, Hokkaido University, Sapporo 060-0810, Japan}
\author{M.~Watanabe, J.~Yamada, and S.~Nakatsuji}
\address{Department of Material Science, Himeji Institute of Technology, Kamigohri 678-1297, Japan}
\date{}
\maketitle
\begin{abstract}
Magnetoresistance measurements have been carried out along the highly conducting $a$ axis
in the FISDW phase of hydrogened and deuterated (TMTSF)$_2$ClO$_4$ for various cooling rates through the anion ordering temperature.
With increasing the cooling rate, 
a) the high field phase boundary $\beta_{\rm {HI}}$, observed at 27 T in hydrogened samples for slowly cooled, 
is shifted towards a lower field, b) the last semimetallic SDW phase below $\beta_{\rm {HI}}$ is suppressed,
and c) the FISDW insulating phase above $\beta_{\rm {HI}}$ is enhanced in both salts.
The cooling rate dependence of the FISDW transition and of $\beta_{\rm {HI}}$ in both salts can be explained by taking into 
account the peculiar SDW nesting vector 
stabilized by the dimerized gap due to anion ordering.

\end{abstract}
%Do not delete this line
} 
\narrowtext 

\section{INTRODUCTION}

The quasi-one-dimensional (Q1D) organic compounds (TMTSF)$_2X$, where TMTSF denotes tetra\-methyl\-tetra\-selena\-fulvalene 
and $X$=PF$_6$, AsF$_6$, ClO$_4$, etc., show many interesting phenomena such as superconductivity, anion ordering (AO), 
spin-density wave (SDW), a cascade of field-induced SDW (FISDW).~\cite{Ishiguro}
The phase diagram of the FISDW phase in the PF$_6$ salt, 
with quantized Hall resistance $\rho_{xy} \sim h/(n2e^2)$ 
in the sequence n=.....4,3,2,1,0 as the magnetic field is increased,
is successfully explained by the mean field theory named the ``standard model" 
based on the nesting of a pair of slightly warped parallel sheets of the Q1D Fermi surface.
The states labeled with integer $n$ have been identified as semimetallic FISDW states
while that with $n$=0 is a FISDW insulating state.
However, the phase diagram of the FISDW phase in hydrogened (TMTSF-h$_{12}$)$_2$ClO$_4$ 
(abbreviated to $h$-ClO$_4$ hereafter)
for the case of slow cooling is known to show disagreements with the ``standard model" i.e.,
(1) in the low-field cascade of FISDW transitions, the sequence of Hall plateaus
is not in the expected order.~\cite{Ribault}
(2) a very stable quantum Hall
state is observed from 7.5 to 27 T.(Ref.\onlinecite{Chamberlin,Noughton})
(3) the FISDW transition temperature $T_{\rm {FISDW}}$
($\sim$ 5.5 K) is independent of field above 15 T.(Ref.\onlinecite{McKernan})
In order to explain the differences between  PF$_6$ and ClO$_4$ salts, 
attention has been focused on the anion ordering which occurs in $h$-ClO$_4$ at 24 K 
and dimerizes the system along the $b$ direction
by a superlattice potential $V$ with a wave vector $Q$=(0,1/2,0).~\cite{Pouget} 
This dimerization separates the original Fermi surface (FS) into two pairs of open orbit FS sheets. 
Although the FISDW phase diagram in the slowly cooled $h$-ClO$_4$ salt
is widely investigated experimentally 
and theoretically, however, it is still in question.

When the $h$-ClO$_4$ salt is rapidly cooled through the anion ordering temperature $T_{\rm {AO}}$, 
the orientations of anions are frozen in two directions at random probability and the SDW phase is induced.
In general, the deuteration of the TMTSF salt is thought to work as a positive chemical pressure 
in the crystal.~\cite{Sinzger}
When the positive chemical pressure by deuteration is applied to the ClO$_4$ salt, 
the nesting of the Q1D Fermi surface becomes more imperfect
and the stabilization of the SDW phase in the intermediate cooled states is suppressed.
As a result, the deuterated (TMTSF-d$_{12}$)$_2$ClO$_4$ salt (abbreviated to $d$-ClO$_4$ hereafter) 
is expected to show the FISDW phase 
in a broad range of cooling rates in contrast with the case of hydrogened ones.

In this paper, we describe the 
cooling rate $\Re_{\rm C}$ dependence of the FISDW phase diagram for the hydrogened, for the smallest cooling rate accessible: 0.0009 K/s, 
and deuterated, for different cooling rates: 0.0009 K/s, 0.018 K/s, 0.67 K/s, ClO$_4$ salts 
under strong magnetic fields applied parallel to the $c^{\ast}$ axis; we
discuss the role of the dimerized gap due to AO controlled by the cooling rate, and that of the chemical pressure
by deuteration in the FISDW phase diagram and we compare the results in deuterated salts with 
those measured in hydrogened ones.

\begin{figure}
\epsfxsize=3.0 in \center \epsfbox{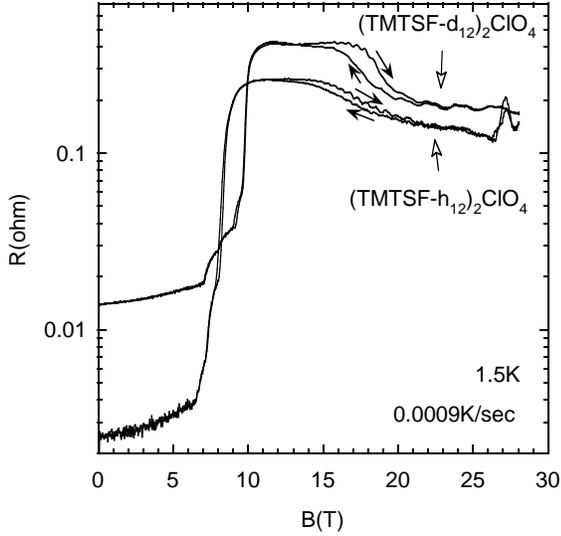} \vspace{0.1in}
\caption{Magnetoresistance along the highly conducting a-axis in hydrogened and deuterated 
(TMTSF)$_2$ClO$_4$ for the relaxed state (cooling rate about 0.0009 K/s)
with the magnetic field parallel to the lowest conductivity direction $c^{\ast}$ at 1.5 K.
} \label{MR-hd}
\end{figure}

\section{Experiments}

Single crystals of (TMTSF)$_2$ClO$_4$ were synthesized by the standard electrochemical method.
The resistance measurements along the conducting $a$ axis
were carried out using a standard four probe dc method  
over the temperature range from 1.5 K to 10 K. 
The typical size of the sample was 1$\times$0.1$\times$0.1 mm$^3$.
Electric leads of 10 $\mu$m gold wire were attached with silver paint onto 
gold evaporated contacts.
The current contacts covered the whole areas of both ends of the sample for an uniform current.
In order to prepare states with various degrees of anion ordering, 
the sample was heated up to 40 K,
and then cooled again with a progressive decrease of the heating to give a controlled cooling rate $\Re_{\rm C}$.
The temperature was measured using a Cernox CX-1050-SD
resistance thermometer calibrated by a capacitance sensor in magnetic fields.
The measurements in the fields to 28 T were done 
in a resistive magnet at the Grenoble High Magnetic Field Laboratory.

\section{RESULTS AND Discussion}

\begin{figure}
\epsfxsize=3.0 in \center \epsfbox{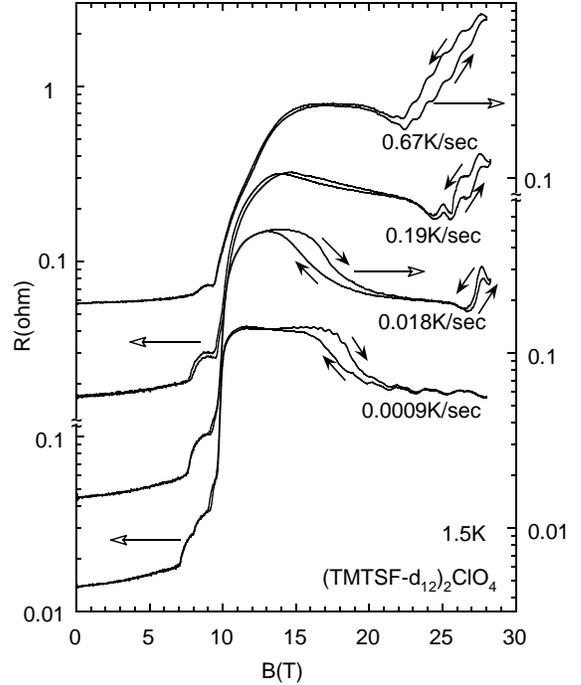} \vspace{0.1in}
\caption{Cooling rate dependence of the magnetoresistance of (TMTSF-d$_{12}$)$_2$ClO$_4$
with the magnetic field parallel to the $c^{\ast}$ direction at 1.5 K.} \label{CRMR}
\end{figure}

Figure \ref{MR-hd} shows the magnetoresistance along the highly conducting $a$ axis in hydrogened and deuterated 
(TMTSF)$_2$ClO$_4$ at 1.5 K for relaxed state, in which $\Re_{\rm C}$ is about 0.0009K/s.
Magnetic field $B$ up to 28 T was applied parallel to the lowest conductivity direction $c^{\ast}$.
For hydrogened $h$-ClO$_4$, it is found from the sudden increase of resistance that,
the transitions to the first and last (n=1) semimetallic SDW phase take place at
about 6.5 T and 8 T, respectively.
With increasing $B$, the nonoscillatory background resistance $R_0$ goes up and has a broad peak between 10 T and 
15 T and a decrease above 15 T.
The magnetoresistance shows hysteresis between 14 and 21 T.
The rapid oscillation (RO) is clearly seen above 14 T in Fig.\ref{MR-hd}.
Above 27 T the field which was proposed by McKernan {\it et al.} as 
a new phase boundary $\beta_{\rm {HI}}$ of a first-order transition \cite{McKernan}, 
both $R_0$ and the amplitude of RO suddenly increases; the 
magnetoresistance shows an isosceles triangle shape.
The sudden increase of $R_0$ and of the RO amplitude is consistent with previous results.~\cite{Brooks}
On the other hand, by deuteration of the ClO$_4$ salt, $T_{\rm {AO}}$ increases from 24 K to 27 K,
and the transition field to the first and last semimetallic SDW phase
increases from 6.5 T to 7 T and from 8 T to 9.7 T, respectively.
In addition, the field of the broad peak shifts to the high field side
and the phase boundary at high field $\beta_{\rm {HI}}$ is not observed below 28 T.
This indicates that deuteration of the ClO$_4$ salt moves the FISDW phase boundary towards the high pressure side
and $\beta_{\rm {HI}}$ is pushed out above 28 T.
These results are regarded as the consequence of a positive chemical pressure 
in the crystal by deuteration.~\cite{Guo,Kang} 
This is consistent with usual deuteration effects.~\cite{Sinzger,Schwenk}
Moreover, the magnetoresistance of $d$-ClO$_4$ shows a 
step-like change 
from the phase between 10 and 17 T to the phase above 20 T with hysteresis between 14 and 21 T.
Although the possibility of a new phase boundary has been proposed around 17 T for $h$-ClO$_4$
from magnetoresistance measurements \cite{Chung}, 
the origin of the above step-like magnetoresistance and hysteresis is unsolved.

The cooling rate $\Re_{\rm C}$ dependence of the magnetoresistance of $d$-ClO$_4$ at 1.5 K is shown in Fig.\ref{CRMR}. 
From this figure, we find that the transition field to the last semimetallic SDW at 9.7 T 
is not sensitive to $\Re_{\rm C}$.
This means that $\Re_{\rm C}$ in this region dose not change the effective pressure in the crystal. 
With increasing $\Re_{\rm C}$, the sudden increase of the resistance at 9.7 T becomes rounded,
and both the RO and the large hysteresis of the magnetoresistance becomes 
dim in the semimetallic SDW phase.
Moreover, $\beta_{\rm {HI}}$ determined from the sudden increase of 
$R_0$ and of the RO amplitude is clearly visible and it is shifted 
towards a lower field with a large hysteresis in the magnetoresistance
when $\Re_{\rm C}$ is increased. 
The increase of the hysteresis loop above $\beta_{\rm {HI}}$ with increasing $\Re_{\rm C}$
is a consequence of the broadening of the first-order transition by disorder effects.
We have also measured the $\Re_{\rm C}$ dependence of the magnetoresistance of $h$-ClO$_4$ and 
observed that $\beta_{\rm {HI}}$ decreases from 27 T for 0.0009 K/s to 21.5 T for 0.018 K/s, 
although $\beta_{\rm {HI}}$ becomes not very well defined for the cooling rate above 0.19 K/s.
One can conclude that $\Re_{\rm C}$ dependence of $\beta_{\rm {HI}}$ is a common feature in ClO$_4$ salts.

\begin{figure}
\epsfxsize=3.0 in \center \epsfbox{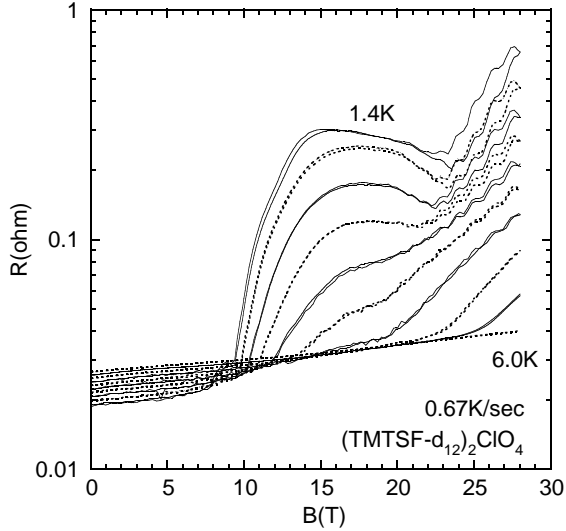} \vspace{0.1in}
\caption{Magnetoresistance of (TMTSF-d$_{12}$)$_2$ClO$_4$ 
with the field parallel to the  $c^{\ast}$ direction
for the cooling rate about 0.67 K/s at 1.3, 2.0, 2.5, 3.0, 3.5, 4.0, 4.3, 5.0, 5.5, and 6.0 K. 
 } \label{0.67}
\end{figure}

\begin{figure}
\epsfxsize=3.0 in \center \epsfbox{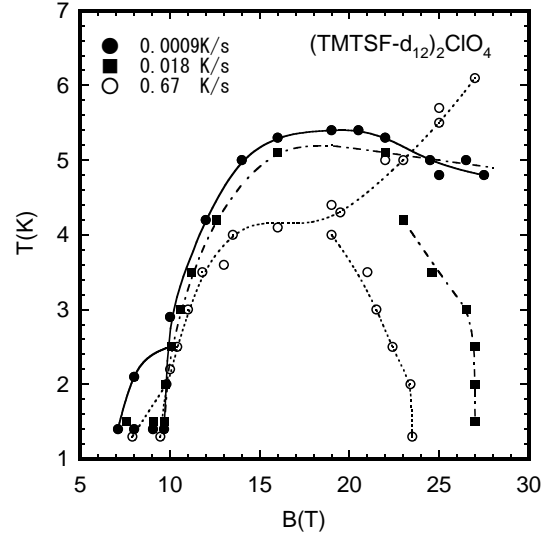} \vspace{0.1in}
\caption{The FISDW phase diagram in deuterated (TMTSF)$_2$ClO$_4$ constructed from many temperature and field 
sweeps for various cooling rates. 
The bold, dashed-and-dotted and dashed lines are guides to the eye
for 0.0009 K/s, 0.018 K/s, and 0.67 K/s cooling rates, respectively. 
 } \label{FD-d}
\end{figure}

\begin{figure}
\epsfxsize=3.0 in \center \epsfbox{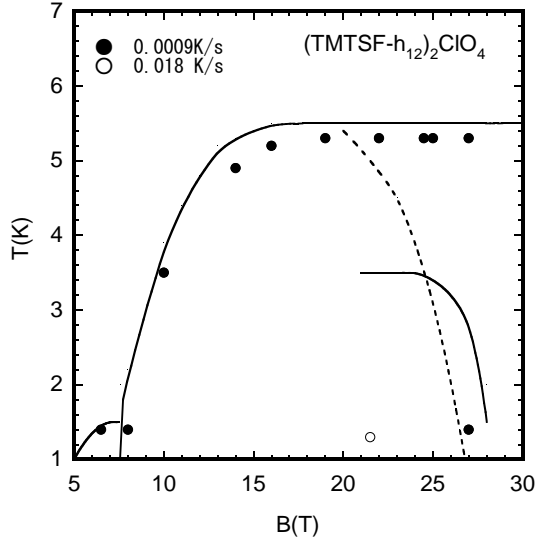} \vspace{0.1in}
\caption{The FISDW phase diagram in hydrogened (TMTSF)$_2$ClO$_4$ 
with the magnetic field parallel to the lowest conductivity direction $c^{\ast}$.
Solid lines show the phase boundary for the relaxed state proposed by McKernan {\it et al.} \cite {McKernan}
The dashed line is the phase boundary of the final FISDW phase for the relaxed state originally determined by Naughton {\it et al.} \cite{Noughton}
Solid and open circles indicate the FISDW transition temperature and the high-field phase boundary for our hydrogened samples.
} \label{FD-h}
\end{figure}

In Fig.\ref{0.67}, we show the magnetoresistance of $d$-ClO$_4$ at various temperatures 
for $\Re_{\rm C}$ about 0.67 K/s.
At this cooling rate, for each temperature, we have determined the value of the FISDW transition 
of $d$-ClO$_4$ as the intersection point between the extrapolations of the magnetoresistance curve 
in the FISDW and metallic phases.
The FISDW transition temperature is plotted as a function of the magnetic field in Fig.\ref{FD-d}. 
As shown in the upper right side of Fig.\ref{FD-d}, it increases with increasing the field $B$.
This result is in agreement with previous measurements on $h$-ClO$_4$
for $\Re_{\rm C}$ = 0.5 K/s.~(Ref.\onlinecite{Matsunaga})
We have previously reported the quadratic increase of $T_{\rm {FISDW}}$ with 
magnetic field above 12 T in $h$-ClO$_4$ for $\Re_{\rm C}$ of 0.5 K/s.~(Ref.\onlinecite{Matsunaga})
In the case of $d$-ClO$_4$, $T_{\rm {FISDW}}$
for $\Re_{\rm C}$ of 0.67 K/s also roughly shows
the quadratic field dependence above 13 T.
However, for intermediate cooling rates, the quadratic field dependence of $T_{\rm {FISDW}}$ 
is not a quadratic increase as predicted by the mean field theory \cite{Montambaux} but 
an envelope line of FISDW states with different quantum numbers $n$.
We have also determined the value of the transition fields at $\beta_{\rm {HI}}$ 
from the intersection of the extrapolations of $R_0$ and have plotted them in the lower right side of Fig.\ref{FD-d}.
As shown in Fig.\ref{0.67}, the large hysteresis in the magnetoresistance and RO are
observed in the high field phase region only below 3 K.
Above 3 K, although the sudden increase of $R_0$ becomes rounded, the hump structure of $R_0$  
exists still up to 4 K.
This means that $\beta_{\rm {HI}}$ exists up to 4 K.
The definition of our phase boundary is the same one determined by Naughton {\it et al.} \cite{Noughton}
as shown in Fig.\ref{FD-h}.
Below the magnetic field of this boundary, the non-zero Hall voltage was observed in all experiments.~\cite{Noughton,McKernan}

As a result, we show in Fig.\ref{FD-h} and Fig.\ref{FD-d} the FISDW phase diagram of $h$-ClO$_4$ and $d$-ClO$_4$ determined 
from magnetoresistance measurements for various $\Re_{\rm C}$, respectively.
In Fig.\ref{FD-h}, solid lines show the phase boundary for the relaxed state
proposed by McKernan {\it et al.}\cite{McKernan}
The dashed line is the phase boundary of the final FISDW phase for the relaxed state originally determined by Naughton {\it et al.} \cite{Noughton}
As shown in Fig.\ref{FD-h}, it is well known that $T_{\rm {FISDW}}$ for a slow cooling in $h$-ClO$_4$
is about 5.5 K and is almost independent of field $B$ above 15 T. 
We confirm this behavior for our hydrogened samples as shown in Fig.\ref{FD-h}.
On the other hand $T_{\rm {FISDW}}$ in $d$-ClO$_4$ 
for $\Re_{\rm C}$ = 0.0009 K/s rapidly increases with increasing $B$ 
up to 16 T and slightly decreases above 20 T as shown in Fig.\ref{FD-d}.
The different behaviour of $T_{\rm {FISDW}}$ at high magnetic field between deuterated samples 
and hydrogened ones  
is attributed to the difference of the chemical pressure in the crystal.
It indicates that the negligible $B$ dependence of $T_{\rm {FISDW}}$ above 15T in hydrogened samples is not intrinsic for 
slowly cooled (TMTSF)$_2$ClO$_4$.
In $d$-ClO$_4$ for 0.018 K/s,
the $B$ dependence of $T_{\rm {FISDW}}$ is almost the same as that for 0.0009 K/s.
We observe, however, that $\beta_{\rm {HI}}$ (27 T) which is independent of temperature below 
2.5 K decreases with increasing temperature above 2.5 K.
Above 3.5 K, $T_{\rm {FISDW}}$ has been determined from the hump structure of $R_0$ which exists up to 4.2 K, although the sudden increase of $R_0$ becomes rounded.
In $d$-ClO$_4$ for 0.67 K/s,
$\beta_{\rm {HI}}$ is observed at about 23.5 T and
$T_{\rm {FISDW}}$ increases with increasing $B$.
The last semimetallic SDW phase between 9.7 T and $\beta_{\rm {HI}}$
 is reduced from 5.5 K to 4 K when the cooling rate $\Re_{\rm C}$ is increased.
This result is consistent with the previous report \cite{Matsunaga} for $h$-ClO$_4$,
the interpretation of which will be discussed later.
Thus, the experimental results lead to the conclusions that, 
with increasing the cooling rate $\Re_{\rm C}$, (1) the high field phase boundary $\beta_{\rm {HI}}$ shifts 
towards a lower field,
(2) the last semimetallic FISDW phase between 9.7 T and $\beta_{\rm {HI}}$ is suppressed,
(3) the FISDW insulating phase above $\beta_{\rm {HI}}$ in which the
Hall voltage becomes almost zero~\cite{McKernan} is enhanced.
These results mean that the last semimetallic FISDW phase and the FISDW insulating 
phase correspond to \underline{different FISDW states}.

\begin{figure}
%\vspace*{10cm}
\epsfxsize=3.0 in \center \epsfbox{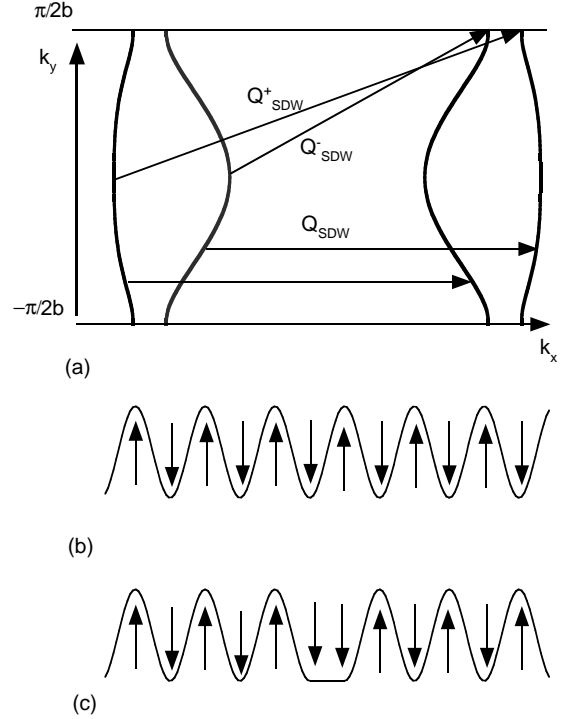} \vspace{0.1in}
\caption{(a) Schematic of the Fermi surfaces of (TMTSF)$_2$ClO$_4$, resulting from a dimerization of 
the system along the $b$ axis. $Q_{\rm {SDW}}$ = (2k$_F$, $\pi$/b) = (2k$_F$, 0) and $Q_{\rm {SDW}}^\pm$ = (2k$_F^\pm$, $\pi$/2b)
are the SDW nesting vectors.
(b) Periodic anion potential without the boundary between anion ordered states.
(c) Anion potential with the boundary between anion ordered states.
} \label{SFS-abc}
\end{figure}

In order to explain this $\Re_{\rm C}$ dependence of the FISDW phase,
we will now discuss the role of the dimerized gap due to anion ordering (AO) and 
that of the SDW nesting vector.
As discussed in a previous report~\cite{Matsunaga},
the concentration of scattering centers associated with 
the boundaries between anion-ordered regions increases with increasing $\Re_{\rm C}$.
This is consistent with the fact that the residual resistance in the metallic 
phase increases with increasing $\Re_{\rm C}$.
For the slowly cooled ClO$_4$ salt, AO creates a superlattice 
potential (Fig.\ref{SFS-abc}(b)) dividing the original Fermi surface into two pairs of open sheets. 
As a result, a dimerized gap due to AO is introduced in the electron band as shown in Fig.\ref{SFS-abc}(a).
Because the periodic anion potential is out of phase at the boundary between adjacent anion-ordered regions (Fig.\ref{SFS-abc}(c)), 
for electrons moving across these boundaries, the dimerized gap due to AO is averaged out.
These boundaries not only work as scattering centers but also suppress the dimerized gap due to AO.
As a result, the effective dimerized gap due to AO decreases with
increasing $\Re_{\rm C}$.
Perturbative calculations using the ``standard model" have shown  
that in the case of a small superlattice potential $V$ due to AO,
the most stable SDW nesting vector $Q_{\rm {SDW}}$ is (2k$_F$, $\pi/b$) = 
(2k$_F$, 0) (see Fig.\ref{SFS-abc}(a)) and subphases with odd quantum numbers, i.e., ... 5, 3, 1,
successively appear when the field is increased.~\cite{Lebed,Osada}
Although this model can explain the phase diagram of slowly cooled 
$h$-ClO$_4$ below 8T(Ref. \onlinecite{Ribault})
and the enhancement of the FISDW insulating phase above the high field phase 
boundary $\beta_{\rm {HI}}$ with increasing $\Re_{\rm C}$,
the decrease of $\beta_{\rm {HI}}$ and the suppression of the last 
semimetallic SDW phase with increasing $\Re_{\rm C}$ can not be explained within the ``standard model" with small $V$.

On the other hand, McKerman {\it et al.} proposed the nesting of another pair of the two pairs 
of open orbit Fermi surface sheets separated by the superlattice potential.~\cite{McKernan}
Using non-perturbative calculations Kishigi claimed that a new FISDW phase with 
a SDW nesting vector $Q_{\rm {SDW}}^\pm$ = (2k$_F^\pm$, $\pi/2b$) 
as shown in Fig.\ref{SFS-abc}(a) is stabilized for a large $V$.~(Ref.\onlinecite{Kishigi})
In fact, the magnitude of $V$ estimated from the angular dependent magnetoresistance measurements is of the order of 
the interchain hopping integral t$_b$.~(Ref.\onlinecite{Yoshino})
Recent calculations for ClO$_4$ salt have pointed out that, 
when $V$ is increased, 
the SDW state with $Q_{\rm {SDW}}$ is rapidly suppressed 
while the SDW state with $Q_{\rm {SDW}}^\pm$ becomes stable at higher value of $V$.~(Ref.\onlinecite{Zanchi,Sengupta})
Because the FISDW phase between 12 and 24 T is suppressed with increasing 
$\Re_{\rm C}$,
it is reasonable to estimate that the respective ground states of the last semimetallic 
FISDW phase and of the insulating FISDW phase 
are a n=1 state with $Q_{\rm {SDW}}^\pm$ and a n=0 insulating state 
with $Q_{\rm {SDW}}$.
Because the FISDW phase with $Q_{\rm {SDW}}^\pm$ becomes more stable with increasing the value of $V$,
the model with $Q_{\rm {SDW}}^\pm$ can explain the decrease 
of $\beta_{\rm {HI}}$ and the 
suppression of the last semimetallic SDW phase.
We are therefore led to conclude that $\beta_{\rm {HI}}$ of our experiment 
corresponds to the phase boundary
between $Q_{\rm {SDW}}^\pm$ phase and $Q_{\rm {SDW}}$ phase.
As shown in Fig.\ref{FD-h}, the high field phase boundary $\beta_{\rm {HI}}$ (a first-order transition) proposed by McKernan {\it et al.} 
is located at 3.5 K between 21 T and 25 T.~(Ref.\onlinecite{McKernan})
Their phase boundary is characterized by the step of Hall voltage 
and magnetization.
They considered this phase boundary as the FISDW transition of 
one pair of Fermi surface within the $Q_{\rm {SDW}}^\pm$ phase.
As a result, the FISDW state at high magnetic fields can be separated three phases
 characterized by Hall voltage, magnetization and the hump structure of $R_0$.
This FISDW phase diagram is consistent with that calculated by Kishigi.~\cite{Kishigi}
Accordingly, if we assume that the last semimetallic FISDW phase is the FISDW phase with  
$Q_{\rm {SDW}}^\pm$ stabilized by the dimerized gap due to AO,
the $\Re_{\rm C}$ dependence of the FISDW transition and of $\beta_{\rm {HI}}$
in the hydrogened and deuterated ClO$_4$ salts can be explained 
by the effective dimerized gap resulting from AO.

\section{Conclusion}

We have measured the magnetoresistance, up to 28 T, in the SDW phase
of hydrogened and deuterated (TMTSF)$_2$ClO$_4$ for various cooling rates $\Re_{\rm C}$ 
through the anion ordering temperature. We thus have
obtained the $\Re_{\rm C}$ dependence of the FISDW transition temperature $T_{\rm {FISDW}}$ and 
of the phase boundary at high magnetic fields.
The deuteration of the ClO$_4$ salt works as a positive chemical pressure in the crystal and
moves the phase boundaries of FISDW towards the high field side.
The deuterated (TMTSF-d$_{12}$)$_2$ClO$_4$ salt shows the FISDW phase in the 
wide cooling rate region which allowed us to investigate the $\Re_{\rm C}$ dependence of the FISDW phase in detail. 
We have found that, with increasing $\Re_{\rm C}$, the high field phase 
boundary $\beta_{\rm {HI}}$ is shifted towards a lower magnetic field,
the last semimetallic SDW phase below $\beta_{\rm {HI}}$ is suppressed,
and the FISDW insulating phase above $\beta_{\rm {HI}}$ is enhanced.
The $\Re_{\rm C}$ dependence of $T_{\rm {FISDW}}$ and of $\beta_{\rm {HI}}$ can 
be explained by the mean-field theory by taking into account the FISDW phase with the SDW nesting vector 
$Q_{\rm {SDW}}^\pm$ stabilized by the dimerized gap due to anion ordering.

\acknowledgments
The authors wish to acknowledge helpful discussions with 
Dr.~K.~Kishigi.
Some of this work was carried out as part of the
``Research for the Future'' project, JSPS-RFTF97P00105,
supported by the Japan Society for the Promotion 
of Science.

\end{document}